\documentclass{elsart}
\usepackage{epsfig}
\usepackage{amsmath}
\usepackage{color}
\usepackage{amssymb}
\usepackage{graphicx}
\usepackage{subfigure}
\usepackage{rotating}
\usepackage{setspace}

\newcommand \be{\begin{equation}}
\newcommand \ba{\begin{eqnarray}}
\newcommand \ee{\end{equation}}
\newcommand \ea{\end{eqnarray}}

\begin{document}

\begin{frontmatter}

\title{Inverse statistics in stock markets: Universality and idiosyncracy}

\author[SKLCRE]{Wei-Xing Zhou\thanksref{EM}},
\author[SKLCRE]{Wei-Kang Yuan}
\address[SKLCRE]{State Key Laboratory of Chemical Reaction
Engineering,\\
East China University of Science and Technology, Shanghai 200237,
China}
\thanks[EM]{Corresponding author. 130 Meilong Road, East China University of Science
and Technology, P.O. Box 369, Shanghai 200237, China\\ {\it E-mail
address:} wxzhou@moho.ess.ucla.edu (W.-X. Zhou)}

\begin{abstract}
Investigations of inverse statistics (a concept borrowed from
turbulence) in stock markets, exemplified with filtered Dow Jones
Industrial Average, S\&P 500, and NASDAQ, have uncovered a novel
stylized fact that the distribution of exit time follows a power
law $p(\tau_\rho) \sim \tau_\rho^{-\alpha}$ with $\alpha \approx
1.5$ at large $\tau_\rho$ and the optimal investment horizon
$\tau_\rho^*$ scales as $\rho^\gamma$
\cite{Simonsen-Jensen-Johansen-2002-EPJB,Jensen-Johansen-Simonsen-2003-PA,Jensen-Johansen-Simonsen-2003-IJMPC}.
We have performed an extensive analysis based on unfiltered daily
indices and stock prices and high-frequency (5-min) records as
well in the markets all over the world. Our analysis confirms that
the power-law distribution of the exit time with an exponent of
about $\alpha=1.5$ is universal for all the data sets analyzed. In
addition, all data sets show that the power-law scaling in the
optimal investment horizon holds, but with idiosyncratic exponent.
Specifically, $\gamma \approx 1.5$ for the daily data in most of
the developed stock markets and the five-minute high-frequency
data, while the $\gamma$ values of the daily indexes and stock
prices in emerging markets are significantly less than 1.5. We
show that there is of little chance that this discrepancy in
$\gamma$ stems from the difference of record sizes in the two
kinds of stock markets.
\end{abstract}


\end{frontmatter}

\section{Introduction}

Econophysics is an interdisciplinary science which applies
statistical physics and complex system theories to economics
\cite{Mandelbrot-1963-JB,Mantegna-Stanley-2000,Bouchaud-Potters-2000,Sornette-2003}.
More than ten stylized facts of asset returns have been discovered
or re-discovered in the community \cite{Cont-2001-QF}, some of
which are inspired originally by the analogy between finance
markets and turbulence
\cite{Ghashghaie-Breymann-Peinke-Talkner-Dodge-1996-Nature,Mantegna-Stanley-1996-Nature}.
In these works, the asset return, as the counterpart of velocity
difference in turbulence, plays a central role, which is defined
as the difference of logarithmic prices at a given time lag.
Recently, a new stylized fact have been unveiled dealing with the
inverse statistics of the exit time in the Dow Jones Industrial
Average
\cite{Simonsen-Jensen-Johansen-2002-EPJB,Jensen-Johansen-Simonsen-2003-PA,Jensen-Johansen-Simonsen-2003-IJMPC}
and in the foreign exchange markets
\cite{Jensen-Johansen-Petroni-Simonsen-2004-XXX}. Interestingly,
this concept of inverse statistics was also borrowed from
turbulence \cite{Jensen-1999-PRL} and applied in turbulence
extensively
\cite{Biferal-Cencini-Vergni-Vulpiani-1999-PRE,Abel-Biferal-Cencini-Falcioni-Vergni-Vulpiani-2000-PD,Biferal-Cencini-Lanotte-Vergni-Vulpiani-2001-PRL,Biferal-Cencini-Lanotte-Vergni-2003-PF,Beaulac-Mydlarski-2004-PF,Roux-Jensen-2004-PRE}.

For a given series of log prices $\{s_i\}$ where $i$ corresponds
to trading days, the exit time (or first passage time) $\tau$ at
time $i$ for a given return threshold $\rho>0$ is defined as the
minimal time span needed for the difference of log prices exceeds
$\rho$ for the first time. In other words, one says mathematically
\begin{equation}
\tau_\rho = \inf \{j-i: s_j - s_i \ge \rho, j>i \}. \label{Eq:tau}
\end{equation}
We see that $\tau_\rho \ge 1$ is integer. If the stock price
rises, $\tau_\rho = 1$. It is argued that more small $\tau_\rho$
in a period implies a bullish market while large $\tau_\rho$
indicates a lasting bearish market. For fractional Brownian motion
of Hurst exponent $H$, Ding and Yang \cite{Ding-Yang-1995-PRE}
have found that the distribution density $p(\tau_\rho)$ scales as
\begin{equation}
p(\tau_\rho) \sim \tau_\rho^{-\alpha}~ \label{Eq:p1}
\end{equation}
with $\alpha=2-H$, when $\tau_\rho$ is large. For Brownian motion,
the exit time distribution has been solved analytically
\cite{Rangarajan-Ding-2000-PRE,Rangarajan-Ding-2000-PLA,Rangarajan-Ding-2000-Fractals}
\begin{equation}
p(\tau_\rho) = \frac{\rho}{\sqrt{4\pi
K\tau_\rho^{3}}}\exp[-\rho^2/4K\tau_\rho]~, \label{Eq:p2}
\end{equation}
where $K$ is the generalized diffusion constant. The most probable
exit time $\tau_\rho^*$ (also called the optimal investment
horizon in Finance) is thus scaled as
\begin{equation}
\tau_\rho^* = 2\rho^2/3~. \label{Eq:OIH1}
\end{equation}

The first passage time was studied in physics, biology and
engineering \cite[and references
therein]{Rangarajan-Ding-2000-PRE,Rangarajan-Ding-2000-PLA,Rangarajan-Ding-2000-Fractals}.
When applying to financial markets, the probability distributions
of the exit time of the filtered DJIA follow a power law as
expressed by Eq.~\ref{Eq:p1} with $\alpha=1.5$
\cite{Simonsen-Jensen-Johansen-2002-EPJB,Jensen-Johansen-Simonsen-2003-PA,Jensen-Johansen-Simonsen-2003-IJMPC}
and the foreign exchange rates show $\alpha=2.4$
\cite{Jensen-Johansen-Petroni-Simonsen-2004-XXX}. For the filtered
DJIA, S\&P 500 and NASDAQ, the optimal investment horizon scales
against $\rho$ as
\begin{equation}
\tau_\rho^* \sim \rho^\gamma~, \label{Eq:OIH2}
\end{equation}
with $\gamma=1.8$, remarkably different from $2$
\cite{Simonsen-Jensen-Johansen-2002-EPJB}.

In this paper, we shall perform an extensive investigation of the
inverse statistics in the stock markets. The data sets consist of
daily stock indexes and stock prices from different markets and
high-frequency (five minutes) data as well. We don't filter the
data so that it is more rational to refer to $\rho$ as the return
threshold.

\section{Probability distribution of exit time}
\label{s1:PDF}

\begin{figure}[htb]
\centering
\includegraphics[width=8.5cm]{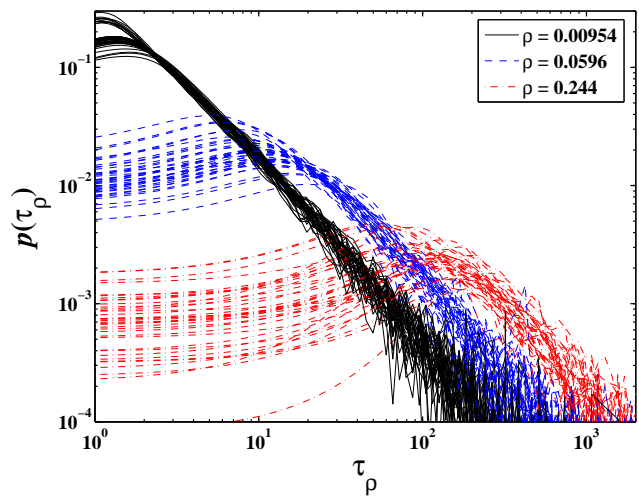}
\caption{\label{Fig:ET:PDF:DJIA} The empirical probability
distributions of the exit time of daily prices of 37 stocks that
are or were components of DJIA since 1962 at three different
levels $\rho=0.00954$, $0.0596$, and $0.244$.}
\end{figure}

The probability distribution of the DJIA has been studied and the
scaling exponent is determined to be $\alpha=1.5$
\cite{Simonsen-Jensen-Johansen-2002-EPJB,Jensen-Johansen-Simonsen-2003-PA}.
To further explore the inverse statistics of the DJIA, we
investigated the daily prices of 37 stocks that are or were
components of DJIA since 1962. The empirical probability densities
$p(\tau_\rho)$ were obtained with the aim of Gaussian kernel
smoothing estimation \cite{Bowman-Azzalini-1997}. The empirical
probability distributions of the exit time for all 37 stocks at
three different levels $\rho=0.00954$, $0.0596$, and $0.244$ are
shown in Fig.~\ref{Fig:ET:PDF:DJIA} with different line types. For
a given $\rho$, the DJIA components have similar shapes in their
probability distributions. This figure also shows the evolution of
the distribution with respect to $\rho$. The power-law scaling is
clearly visible for large $\tau_\rho$. We find that the exponent
of the power law is also $\alpha \approx 1.5$, in agreement with
that of the DJIA.

We also performed analysis on the daily evolution of the S\&P 500
index (from 1940/11/29 to 2004/09/21, totally 16115 data points)
and of the NASDAQ index (from 1971/02/05 to 2002/11/15, totally
8025 data points) in the USA stock markets. The power-law scaling
distribution of the exit time with an exponent of about 1.5 is
confirmed for both indexes, as reported in
\cite{Jensen-Johansen-Simonsen-2003-PA,Jensen-Johansen-Simonsen-2003-IJMPC}.

A question arises naturally asking whether this new statistical
property holds in other stock markets other than the USA. To
address this question, we have carried out the same analysis on
the following 40 indexes all over the world: Argentina (from
1996/10/08 to 2004/09/20), Australia (from 1984/08/03 to
2004/09/21), Austria (from 1986/01/08 to 2002/11/15), Belgium
(from 1991/01/02 to 2002/11/18), Brazil (from 1993/04/27 to
2004/09/20), Canada (from 1982/01/29 to 2002/11/15), Chile (from
1997/06/09 to 2002/10/14), Czech (from 1997/07/01 to 2004/09/20),
Denmark (from 1989/12/04 to 2002/11/15), Egypt (from 1997/07/02 to
2004/09/20), France (from 1987/07/09 to 2002/11/15), Germany (from
1959/10/01 to 2002/11/15), Hong Kong (from 1969/11/24 to
2004/09/21), India (from 1997/07/01 to 2004/09/20), Indonesia
(from 1997/07/01 to 2004/09/20), Ireland (from 1987/03/09 to
2002/11/15), Israel (from 1992/10/08 to 2004/09/20), Italy (from
1992/12/31 to 2002/11/15), Japan (from 1984/01/04 to 2004/09/17),
Korea (from 1997/07/01 to 2004/09/20), Malaysia (from 1977/01/03
to 2002/11/15), Mexico (from 1992/01/02 to 2002/11/15), The
Netherlands (from 1983/01/03 to 2002/11/15), New Zealand (from
1988/08/31 to 2002/11/15), Norway (from 1987/01/02 to 2002/11/15),
Pakistan (from 1997/07/02 to 2004/09/20), Philippines (from
1997/07/02 to 2004/09/20), Russia (from 1997/07/01 to 2004/09/20),
SP500 (from 1940/11/29 to 2004/09/21), Singapore (from 1985/01/04
to 2002/11/15), Slovakia (from 1997/07/01 to 2002/10/14), South
Africa (from 1995/06/30 to 2002/11/15), Spain (from 1987/01/05 to
2002/11/15), Sri Lanka (from 1997/07/01 to 2004/09/20), Sweden
(from 1986/12/18 to 2002/11/15), Switzerland (from 1989/07/03 to
2002/11/15), Taiwan (from 1997/07/02 to 2004/09/20), Thailand
(from 1997/07/02 to 2004/09/20), Turkey (from 1997/07/01 to
2004/09/20), UK (from 1984/04/02 to 2004/09/21), Venezuela (from
2000/07/04 to 2002/09/05). We also observed a power-law
distribution of the exit time with an exponent close to 1.5.

\begin{figure}[htb]
\centering
\includegraphics[width=8.5cm]{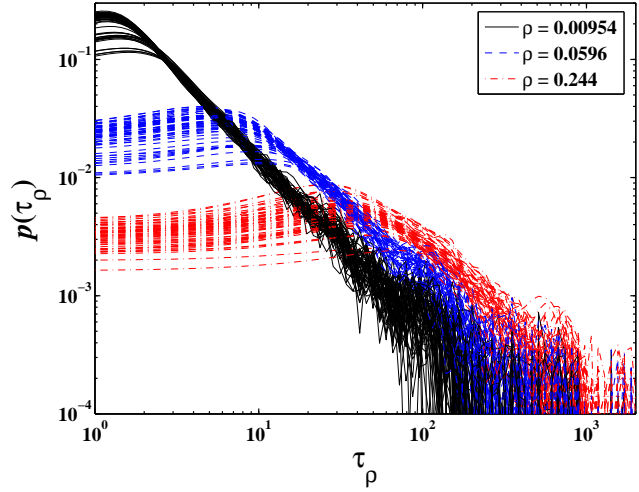}
\caption{\label{Fig:ET:PDF:CN} The empirical probability
distributions of the exit time of daily prices of the Shanghai
Stock Exchange Composite, A Share Index, B Share Index, and 55
stocks listed in Shanghai Stock Exchange and Shenzhen Stock
Exchange before 1992/12/01 at three different levels
$\rho=0.00954$, $0.0596$, and $0.244$.}
\end{figure}

As a typical example of emerging markets, we present in
Fig.~\ref{Fig:ET:PDF:CN} the empirical probability distributions
of the exit time of daily prices of the Shanghai Stock Exchange
Composite, A Share Index, B Share Index, and 55 stocks listed in
Shanghai Stock Exchange and Shenzhen Stock Exchange before
1992/12/01 at three different levels $\rho=0.00954$, $0.0596$, and
$0.244$. One finds again that Eq.~(\ref{Eq:OIH1}) holds and
$\alpha \approx 1.5$.

\begin{figure}[htb]
\centering
\includegraphics[width=8.5cm]{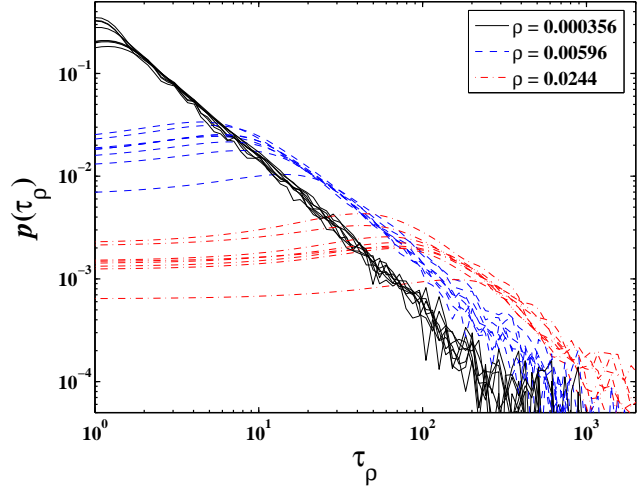}
\caption{\label{Fig:ET:PDF:CN:HF} The empirical probability
distributions of the exit time of high-frequency (5 minutes)
prices of the Shanghai Stock Exchange Composite and 7 stocks from
2002/07/01 (or 2002/10/09, or 2003/07/25) to 2004/09/16 in China
at three different levels $\rho=0.000954$, $0.00596$, and
$0.0244$.}
\end{figure}

So far, we have checked the validity of the new stylized fact in
the daily evolutions of indexes and stock prices in both developed
markets and emerging markets. To further clarify the situation,
explorations were carried out on 10 high-frequency (5 minutes)
data sets, including CAC 40 of France (covering July, August,
September, and October of 2001, totally 9332 data points), S\&P
500 of USA (covering 1995, 1996, and 1997, totally 59885 data
points), SSEC of China (from 2002/07/01 to 2004/09/16, totally
24955 data points), and seven stocks traded in the Shanghai Stock
Exchange (six of them from June 2002, one from 2002/10/09, and one
from 2003/07/25, all up to 2004/09/16). Again, the scaling law
(\ref{Eq:OIH1}) holds and $\alpha \approx 1.5$. Figure
\ref{Fig:ET:PDF:CN:HF} illustrates the distributions of the data
sets in the Chinese stock market.

We also analyzed the 1-minute-by-1-minute intraday NASDAQ data
over 50 open days in the past from Monday Oct. 30, 2000. We
observed a clear crossover from one power-law scaling to another
separated with a kink around $\tau_\rho \approx 80$. The scaling
exponent in the region $\tau_\rho^*<\tau_\rho < 80$ is $\alpha_1 =
1.1$. The scaling exponent in the region $\tau_\rho > 80$ is hard
to estimate due to statistics. However, we can see that $\alpha_2$
of the second region is comparable to 1.5 or ever larger. This may
signal a possible discrepancy between data sets of sampling
frequency higher or lower than 1/min. We shall come back to this
in the future when longer data sets are available.

\section{Optimal investment horizon}
\label{s1:OIH}

In the previous section, we verify the finding of a power-law
scaling with an exponent of about 1.5
\cite{Simonsen-Jensen-Johansen-2002-EPJB,Jensen-Johansen-Simonsen-2003-PA,Jensen-Johansen-Simonsen-2003-IJMPC}
using numerous stock market indexes and individual stocks whose
recording frequency is daily or 5 min all over the world. In this
section, we focus on the power law relation (\ref{Eq:OIH2})
between the optimal investment horizon $\tau_\rho^*$ and the
threshold $\rho$. We find that this power law (\ref{Eq:OIH2})
holds for all the data sets we studied in Section \ref{s1:PDF} but
with quite different exponents.

\begin{figure}[htb]
\centering
\includegraphics[width=8.5cm]{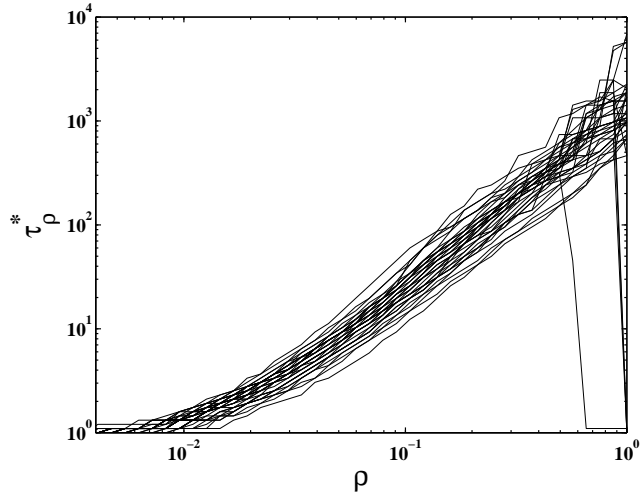}
\caption{\label{Fig:ET:taurho:DJIA} The dependence of optimal
investment horizon $\tau_\rho^*$ with respect to $\rho$ of daily
prices of 37 stocks that are or were components of DJIA since
1962. The three lines with clear cutoffs at large $\rho$
correspond to stock prices with relatively short sizes.}
\end{figure}

For daily DJIA (1896-2004), we find that $\gamma \approx 1.55$,
while S\&P 500 (1940-2004) and NASDAQ (1971-2002) give $\gamma
\approx 1.5$. These values are significantly lower than those
reported in
\cite{Simonsen-Jensen-Johansen-2002-EPJB,Jensen-Johansen-Simonsen-2003-PA,Jensen-Johansen-Simonsen-2003-IJMPC}.
This discrepancy may be due to the fact that we don't preprocess
the data with a high-pass filter. Calculations with the 37 stocks
in DJIA give similar results, as illustrated in
Fig.~\ref{Fig:ET:taurho:DJIA}.

\begin{figure}[htb]
\centering
\includegraphics[width=8.5cm]{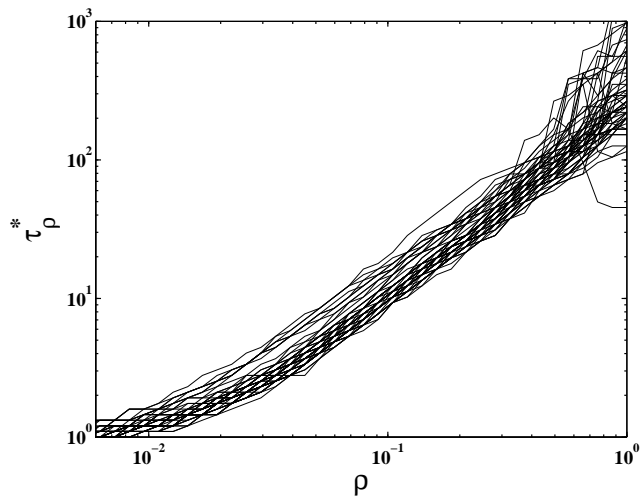}
\caption{\label{Fig:ET:taurho:CN} The dependence of optimal
investment horizon $\tau_\rho^*$ with respect to $\rho$ of daily
prices of the Shanghai Stock Exchange Composite, A Share Index, B
Share Index, and 55 stocks listed in Shanghai Stock Exchange and
Shenzhen Stock Exchange before 1992/12/01.}
\end{figure}

For the 40 indexes in the stock markets throughout the world that
we studied in the previous section, 22 stock markets have an
exponent $\gamma \approx 1.50-1.55$. The rest 18 indexes have
different $\gamma$ values. The Taiwan stock market index gives a
higher scaling exponent $\gamma=1.64$, while the others have
smaller $\gamma$ values (Belgium 1.40, Brazil 1.42, Chile 1.20,
Czech 1.42, Egypt 1.20, Indonesia 1.35, Israel 1.40, Korea 1.39,
Malaysia 1.36, New Zealand 1.40, Pakistan 1.37, Philippines 1.29,
Russia 1.37, Singapore 1.38, Sri Lanka 1.39, Turkey 1.33,
Venezuela 1.0). With only a few exceptions, we see that most of
the 22 stock markets having $\gamma \approx 1.50-1.55$ are
developed markets, while most of 18 remaining indexes are emergent
markets. In addition, Fig.~\ref{Fig:ET:taurho:CN} shows the
dependence on $\rho$ of $\tau_\rho^*$ of 58 time series in the
Chinese stock market. The value of $\gamma$ is around 1.25.
Therefore, the exponent $\gamma$ is a kind of measure of the
maturity of stock market in some sense.

\begin{figure}[htb]
\centering
\includegraphics[width=8.5cm]{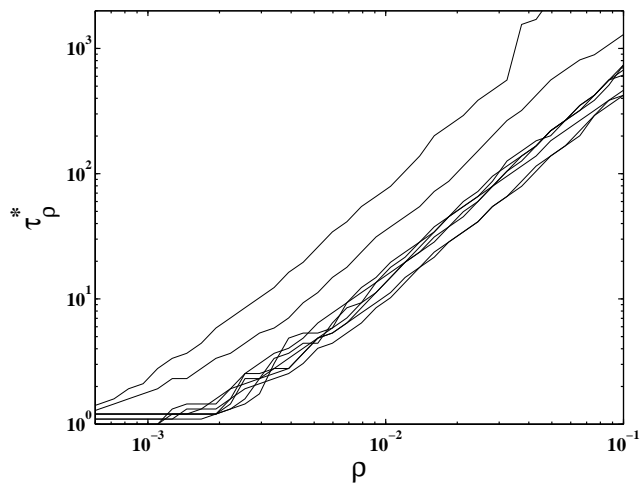}
\caption{\label{Fig:ET:taurho:CN:HF} The dependence of optimal
investment horizon $\tau_\rho^*$ with respect to $\rho$ of
high-frequency (5 minutes) prices of the Shanghai Stock Exchange
Composite and 7 stocks from 2002/07/01 (or 2002/10/09, or
2003/07/25) in China.}
\end{figure}

In Fig.~\ref{Fig:ET:taurho:CN:HF} shows the dependence of optimal
investment horizon $\tau_\rho^*$ with respect to $\rho$ of
high-frequency (5 minutes) prices of the Shanghai Stock Exchange
Composite and 7 stocks in China. It is interesting to notice that
the exponents of the six high-frequency time series in the Chinese
stock markets are close to 1.50. The CAC 40 and S\&P 500
high-frequency data have $\gamma\approx 1.50$ as expected.

\section{Does the size matter?}
\label{s1:Size}

In Section \ref{s1:OIH}, we have seen that the $\gamma$ values of
most of the emerging markets are less than 1.4 and most of indexes
or stock prices with $\gamma<1.4$ are from emerging markets. For
an emerging market, the sizes of its recorded index and stock
prices are usually short. In China, for instance, the first market
for government-approved securities was founded in Shanghai on
November 26, 1990 and started operating on December 19 of the same
year under the name of the Shanghai Stock Exchange, and shortly
after, the Shenzhen Stock Exchange was established on December 1,
1990 and started its operations on July 3, 1991 \cite{Su-2003}. On
the contrary, most of the major western stock markets have much
longer histories. It is thus of great importance to check whether
such a significant difference in scaling behaviors of the optimal
investment horizon is due to the short size of the recorded time
series in the emerging markets.

To understand this idiosyncratic scaling behavior in the emerging
stock markets, we take the daily DJIA index from 1896 to 2004 as a
proxy. This time series has 29493 data points. We run a moving
window of size 3500 (about 14 years) along the evolution of DJIA
with a step of 20 trading days. This gives 1300 moving windows.
For each window, we determine its exit times of all points
{\it{separately}}. In other words, each running window is treated
as an independent time series. Then the optimal in vestment
horizon $\tau_\rho^*$ is determined for each window at 21
threshold levels logarithmically spaced in $[0.0194, 0.3237]$. To
estimate the value of $\gamma$ for each window, we require a
linear regression coefficient larger than 0.995. If this
constraint is not fulfilled, we remove the last point and regress
the remaining data again. This iterative procedure repeats again
and again until the linear regression coefficient is larger than
0.995. For most of the data sets, this constraint satisfies with
all the 21 points.

\begin{figure}[htb]
\centering
\includegraphics[width=8.5cm]{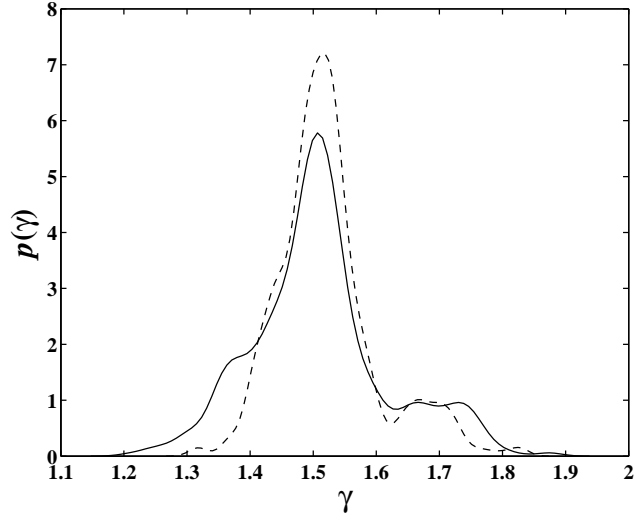}
\caption{\label{Fig:ET:EvoDJIA1} Probability distribution of the
estimated scaling exponent $\gamma$. The probability distributions
are estimated with Gaussian kernel smoothing approach.}
\end{figure}

The probability distribution $p(\gamma)$ of the estimated
$\gamma$'s is plotted in Fig.~\ref{Fig:ET:EvoDJIA1} with solid
line. The maximum locates at $\gamma_{\max}=1.51$. The percentage
of the windows with $\gamma<1.4$ is $13.85\%$. When the Chinese
stock markets are concerned, only 22 windows have $\gamma<1.3$,
implying a percentage of $1.69\%$. This result already supports
the point that there is little chance that the idiosyncratic
scaling behavior in the Chinese stock markets could not be
attributed to its short history and this property is thus of
little probability to be artificial. This conclusion is still
significant although less strong in other emergent markets.

\begin{figure}[htb]
\centering
\includegraphics[width=8.5cm]{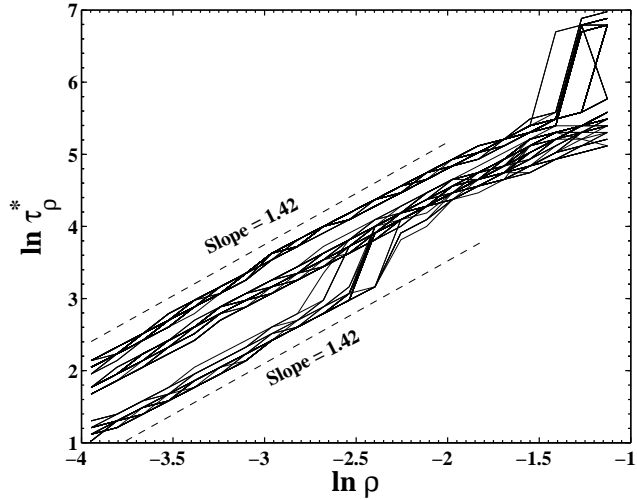}
\caption{\label{Fig:ET:EvoDJIA2} Dependence of $\tau^*_\rho$
against $\rho$ for the 180 windows where $\gamma < 1.4$.}
\end{figure}

To further enhance our point, we have taken a closer look at the
180 regressions where $\gamma<1.4$. Figure \ref{Fig:ET:EvoDJIA2}
shows the dependence of $\tau^*_\rho$ against $\rho$ for all the
180 windows. There are three clusters in
Fig.~\ref{Fig:ET:EvoDJIA2}. The lines in the bottom cluster
exhibit a sudden jump around $\rho=0.08$. The left parts with
relatively small $\rho$ of all lines are linear with a slope of
1.42. The right parts show slightly downward bending. If we fit
the points in the left part for each line, we find that the
percentage of running windows with $\gamma<1.4$ reduces
dramatically to $2.15\%$, while the number of windows with
$\gamma<1.3$ is null.

\begin{figure}[htb]
\centering
\includegraphics[width=8.5cm]{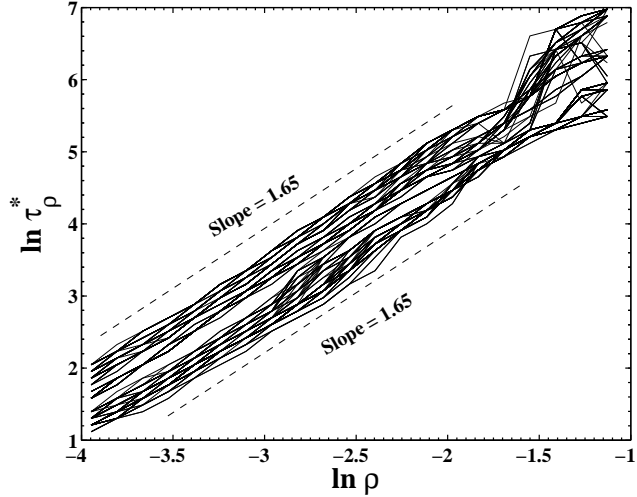}
\caption{\label{Fig:ET:EvoDJIA3} Dependence of $\tau^*_\rho$
against $\rho$ for the 180 windows where $\gamma >1.6$.}
\end{figure}

In Fig.~\ref{Fig:ET:EvoDJIA1}, one observes that there are also
quite a few windows whose $\gamma>1.6$. The number of these
windows is 214, a percentage of $16.46\%$. Figure
\ref{Fig:ET:EvoDJIA3} shows the dependence of $\tau^*_\rho$
against $\rho$ for all the 214 windows. For large $\rho$, the
linearity deteriorates remarkably. The slopes of these lines are
comparable to 1.65. If we redo the regressions after excluding the
points at large $\rho$, the percentage of windows with
$\gamma>1.6$ reduces to $12.31\%$. The probability distribution
$p(\gamma)$ of the modified $\gamma$'s is plotted in
Fig.~\ref{Fig:ET:EvoDJIA1} with dashed line. The maximum locates
at $\gamma_{\max}=1.52$. The kurtosis excess of the distribution
in dashed line is 4.56, higher than that in solid line with a
value of 3.50.

At last, let us average $\tau_\rho^*$ over the 1300 running
windows for each threshold $\rho$. A perfect power law is observed
as follows
\begin{equation}
 \ln \langle\tau_\rho^*\rangle = 1.56 \ln \rho + 7.86 ~.
\label{Eq:AveTau}
\end{equation}
The linear regression coefficient is 0.9997. The power-law
exponent $1.56$ is slightly larger than the $\gamma_{\max}$ due to
the fact that the skewness of the two distributions shown in
Fig.~\ref{Fig:ET:EvoDJIA1} are 0.46 and 0.96, respectively.

\section{Concluding remarks}

There are universal and idiosyncratic behaviors in the developed
and emerging stock markets, such as the log-periodic power-law
antibubble pattern \cite{Zhou-Sornette-2004-PA} or return
anomalies \cite{Mookerjee-Yu-1999-RFE} in China's stock markets.
In this paper, we have verified that the probability density of
exit time follows a universal power law with exponent $\alpha=1.5$
for large $\tau_\rho$ for daily indexes and stock prices and 5-min
high-frequency data in both the developed and emerging markets.
However, this statistical property can not be extrapolated to
other financial markets, at least not to the tick-by-tick data in
the foreign exchange markets where the power law exponent
$\alpha\approx 2.4$ is remarkably greater than the one in stock
markets \cite{Jensen-Johansen-Petroni-Simonsen-2004-XXX}.

On the other hand, although the optimal investment horizon
$\tau_\rho^*$ scales as $\rho^\gamma$ for all stock markets
investigated, they have quite different values of $\gamma$.
Roughly speaking, the $\gamma$ values of the daily data in
emerging stock markets are significantly less than those in major
western markets ($\gamma\approx 1.5$). We have showed that this
discrepancy in $\gamma$ couldn't stems from the difference of
record sizes in the two kinds of stock markets. In words, for a
given gain threshold, the optimal investment horizon in an
emerging market is shorter than that in a developed market. This
is consistent with the known result that the optimal trading time
lag is about 30 days in China's stock markets, much shorter than
that in the USA markets, due to the speculative nature of emerging
markets \cite{Zhou-Sornette-2004-PA}.

For 5-min high-frequency data (CAC 40 index, S\&P 500 index, and
SSEC and stock prices in China) we find that $\gamma\approx 1.5$.
This intriguing feature can be interpreted by the $t+1$ trading
rule in the Chinese stock markets which reduces the speculation at
high-frequency levels. As a cautionary note, we notice that the
1-min high-frequency data of NASDAQ do not follow the simple
power-law relation in the exit time distribution, which calls for
a further investigation.

{\textbf{Acknowledgments:}}

We are grateful to Wei Deng for providing us the numerous data
sets of the indexes and individual stock prices in the Chinese
stock markets and thank Didier Sornette for careful reading of the
manuscript. This work was jointly supported by NSFC/PetroChina
through a major project on multiscale methodology.

\bibliography{ET_Stock_PDF}

\begin{thebibliography}{10}
\expandafter\ifx\csname url\endcsname\relax
  \def\url#1{\texttt{#1}}\fi
\expandafter\ifx\csname urlprefix\endcsname\relax\def\urlprefix{URL }\fi

\bibitem{Simonsen-Jensen-Johansen-2002-EPJB}
I.~Simonsen, M.~H. Jensen, A.~Johansen, Optimal investment horizons, Eur. Phys.
  J. B 27 (2002) 583--586.

\bibitem{Jensen-Johansen-Simonsen-2003-PA}
M.~H. Jensen, A.~Johansen, I.~Simonsen, Inverse statistics in economics: the
  gain-loss asymmetry, Physica A 324 (2003) 338--343.

\bibitem{Jensen-Johansen-Simonsen-2003-IJMPC}
M.~H. Jensen, A.~Johansen, I.~Simonsen, Inverse fractal statistics in
  turbulence and finance, Int. J. Mod. Phys. B 17 (2003) 4003--4012.

\bibitem{Mandelbrot-1963-JB}
B.~B. Mandelbrot, The variation of certain speculative prices, J. Business 36
  (1963) 394--419.

\bibitem{Mantegna-Stanley-2000}
R.~N. Mantegna, H.~E. Stanley, An Introduction to Econophysics: Correlations
  and Complexity in Finance, Cambridge University Press, Cambridge, 2000.

\bibitem{Bouchaud-Potters-2000}
J.-P. Bouchaud, M.~Potters, Theory of Financial Risks: From Statistical Physics
  to Risk Management, Cambridge University Press, Cambridge, 2000.

\bibitem{Sornette-2003}
D.~Sornette, Why Stock Market Crashes: Critical Events in Complex Financial
  Systems, Princeton University Press, Princeton, 2003.

\bibitem{Cont-2001-QF}
R.~Cont, Empirical properties of asset returns: stylized facts and statistical
  issues, Quantitative Finance 1 (2001) 223--236.

\bibitem{Ghashghaie-Breymann-Peinke-Talkner-Dodge-1996-Nature}
S.~Ghashghaie, W.~Breymann, J.~Peinke, P.~Talkner, Y.~Dodge, Turbulent cascades
  in foreign exchange markets, Nature 381 (1996) 767--770.

\bibitem{Mantegna-Stanley-1996-Nature}
R.~N. Mantegna, H.~E. Stanley, Turbulence and financial markets, Nature 383
  (1996) 587--588.

\bibitem{Jensen-Johansen-Petroni-Simonsen-2004-XXX}
M.~H. Jensen, A.~Johansen, F.~Petroni, I.~Simonsen, Inverse statistics in the
  foreign exchange market, Physica A 340 (2004) 678--684.

\bibitem{Jensen-1999-PRL}
M.~H. Jensen, Multiscaling and structure functions in turbulence: An
  alternative approach, Phys. Rev. Lett. 83 (1999) 76--79.

\bibitem{Biferal-Cencini-Vergni-Vulpiani-1999-PRE}
L.~Biferal, M.~Cencini, D.~Vergni, A.~Vulpiani, Exit time of turbulent signals:
  A way to detect the intermediate dissipative range, Phys. Rev. E 60 (1999)
  R6295--R6298.

\bibitem{Abel-Biferal-Cencini-Falcioni-Vergni-Vulpiani-2000-PD}
M.~Abel, L.~Biferal, M.~Cencini, M.~Falcioni, D.~Vergni, A.~Vulpiani,
  Exit-times and $\epsilon$-entropy for dynamical systems, stochastic
  processes, and turbulence, Physica D 147 (2000) 12--35.

\bibitem{Biferal-Cencini-Lanotte-Vergni-Vulpiani-2001-PRL}
L.~Biferal, M.~Cencini, A.~S. Lanotte, D.~Vergni, A.~Vulpiani, Inverse
  statistics of smooth signals: The case of two dimensional turbulence, Phys.
  Rev. Lett. 87 (2001) 124501.

\bibitem{Biferal-Cencini-Lanotte-Vergni-2003-PF}
L.~Biferal, M.~Cencini, A.~S. Lanotte, D.~Vergni, Inverse velocity statistics
  in two-dimensional turbulence, Phys. Fluids 15 (2003) 1012--1020.

\bibitem{Beaulac-Mydlarski-2004-PF}
S.~Beaulac, L.~Mydlarski, Inverse structure functions of temperature in
  grid-generated turbulence, Phys. Fluids 16 (2004) 2126--2129.

\bibitem{Roux-Jensen-2004-PRE}
S.~Roux, M.~H. Jensen, Dual multifractal spectra, Phys. Rev. E 69 (2004)
  016309.

\bibitem{Ding-Yang-1995-PRE}
M.-Z. Ding, W.-M. Yang, Distribution of the first return time in fractional
  brownian motion and its application to the study of on-off intermittency,
  Phys. Rev. E 52 (1995) 207--213.

\bibitem{Rangarajan-Ding-2000-PRE}
G.~Rangarajan, M.-Z. Ding, Anomalous diffusion and the first passage time
  problem, Phys. Rev. E 61 (2000) 120--133.

\bibitem{Rangarajan-Ding-2000-PLA}
G.~Rangarajan, M.-Z. Ding, First passage time distribution for anomalous
  diffusion, Phys. Lett. A 273 (2000) 322--330.

\bibitem{Rangarajan-Ding-2000-Fractals}
G.~Rangarajan, M.-Z. Ding, First passage time problem for biased
  continuous-time random walks, Fractals 8 (2000) 139--145.

\bibitem{Bowman-Azzalini-1997}
A.~Bowman, A.~Azzalini, Applied Smoothing Techniques for Data Analysis, Oxford
  University Press, Oxford, 1997.

\bibitem{Su-2003}
D.-W. Su, Chinese Stock Markets: A Research Handbook, World Scientific,
  Singapore, 2003.

\bibitem{Zhou-Sornette-2004-PA}
W.-X. Zhou, D.~Sornette, Antibubble and prediction of china's stock market and
  real-estate, Physica A 337 (2004) 243--268.

\bibitem{Mookerjee-Yu-1999-RFE}
R.~Mookerjee, Q.~Yu, An empirical analysis of the equity markets in china,
  Review of Financial Economics 8 (1999) 41--60.

\end{thebibliography}

\end{document}